# PHYSICS OF THE ENHANCED OPTICAL COOLING OF PARTICLE BEAMS IN STORAGE RINGS*

E.G.Bessonov[#], Lebedev Phys. Inst. RAS, Moscow, Russia
A.A.Mikhailichenko, Cornell University, Ithaca, NY, U.S.A.
A.V.Poseryaev, Moscow State University

*Abstract*

Physics of enhanced optical cooling (EOC) of particle beams in storage rings, nonlinear features of cooling and requirements to ring lattices, optical and laser systems are discussed.

## INTRODUCTION

The idea of EOC of particle beams in a storage ring is based on external selectivity of interaction between particles and their amplified undulator radiation wavelets (URW) [1], [2]. The scheme of EOC is presented on Figure 1. URW is emitted in a pick-up undulator, focused, amplified in an optical amplifier and pass a kicker undulator together with the same particle. Geometrical parameters of electron and optical beam lines are chosen such a way that particle enters the kicker undulator at decelerating phase. Particles interact with their URWs effectively in the regime of small and moderate current (see below). The selectivity of interaction arranged by a moving screen located in the image plane of optical system projecting URW there.

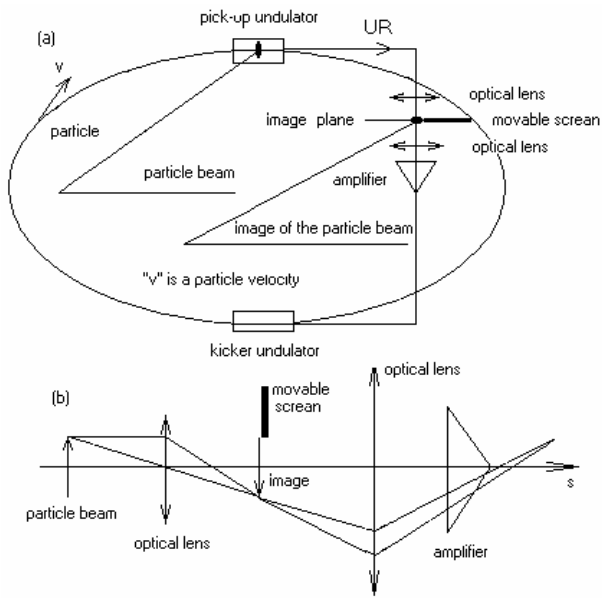

Fig.1. The scheme of EOC of particle beams in a storage ring (a) and unwrapped optical scheme (b).

For any particle in a storage ring, the change of the square of its amplitude of betatron oscillations caused by sudden energy change $\delta E$ in smooth approximation is determined by the equation

___________________
*Supported by RFBR under grant No 05-02-17162 and by NSF.
[#]Corresponding author; bessonov@x4u.lebedev.ru

$$\delta A_x^2 = -2x_\beta \delta x_\eta + (\delta x_\eta)^2, \qquad (1)$$

where $x_\beta$ is the initial particle deviation from it's closed orbit; $\delta x_\eta = \eta_x \beta^{-2}(\delta E/E)$ is the change of it's closed orbit position; $E$ is the energy of the particle; $\eta_x$ is the dispersion function in the storage ring at the location of kicker undulators; $\beta$ is the normalized velocity. In the approximation $|\delta x_\eta| < 2|x_\beta| < 2A_x$, according to (1), the energy loss of a particle $\delta E < 0$ leads to a decrease of its amplitude, if $\eta_x > 0$ and the product

$$x_\beta \delta x_\eta > 0. \qquad (2)$$

Based on this observation, an enhanced scheme of emittance exchange under conditions of week cooling in the frame of the Robinson's damping criterion was sugested and divelloped for unbunched beam (RF turned off) in [1], [3]. Later three EOC schemes of the same principle were suggested in [1], [2]. Below the scheme of EOC is developed.

## PHYSICS OF EOC

Different lattice and optic geometries can be used in EOC. Below we consider some of them.

### Geometry I

Two or more identical undulators are installed in straight sections of a storage ring at a distance determined by a betatron phase advance for the lattice segment $(2p+1)\pi$ between pick-up and first kicker undulator and $2p'\pi$ between next kicker undulators; where $p$, $p' =1,2,3...$ are integer numbers. Untransparent screen is moved out of the image, URWs pass by the screen, amplified and pass through kicker undulators.

Let $\eta_x > 0$ at locations of undulators and screens are absent. Particles enter kicker undulators at decelerating phases of URW. Then if a deviation of a particle in the pick-up undulator $x_\beta > 0$, the deviation of the particle in the kicker undulators comes to be $x_\beta < 0$ and, according to (2), the energy loss by the particle is accompanied by a decrease of both its energy and amplitude of betatron oscillations. On the contrary, if the deviation of the particle in the pick-up undulator $x_\beta < 0$, the deviation of a particle in kicker undulators comes to be $x_\beta > 0$ and, according to (2), the energy loss by the particle is accompanied by a decrease in its energy and increase of its amplitude of betatron oscillations. As a result, the energy spread of the beam is not changed because of all particles lose their energy independent on the closed orbit position, the angular spread, according to (2), is not changed in the first approximation because of all particles lose their energy both at the conditions $x_\beta \delta x_\eta > 0$ and $x_\beta \delta x_\eta < 0$ with equal probability. The beam is displaced only as a whole in the radial direction.

To cool particle beam we have to introduce selectivity of interaction between particles and their URWs. Below we consider physics of EOC in three cooling schemes for the above geometry in case of unbunched particle beams if an optical systems including untransparent screens are used for selection.

I. The edge of a screen is shifted to the initial position of the edge of the particle beam image corresponding to minimum particle energy and stay at rest (the way for URW beam is open). In this case particles loose their energy in the combined fields of amplified URW and kicker undulator, their closed orbits go to the closed orbit of minimum initial energy and the image of the particle beam moves to the screen. As a result, closed orbits of particles will be stopped near to the position corresponding to the minimum energy of particles of the beam after their URW beam will be overlapped by the screen in the image plane and the way for URW beam will be closed.

In this case the position of the closed orbit of a particle, the dispersive beam size $2\sigma_\eta$ and the energy spread of the particle beam $2\sigma_E$ will be changed by the law

$$\partial x_\eta / \partial t = \dot{x}_{\eta,in}, \qquad (3)$$

$$\sigma_\eta = \sigma_{\eta,0}(1 - \frac{\dot{x}_{\eta,in}}{\sigma_{\eta,0}}t), \; \sigma_E = \sigma_{E,0}(1 - \frac{\overline{P}}{\sigma_{E,0}}t) \qquad (4)$$

where $\dot{x}_{\eta,in} = -\eta_x \beta^{-2} (\overline{P}/E)$ is the maximum velocity of closed orbits of particles under condition of absence of screens, $\overline{P}$ is the average rate of particle energy loss in the kicker undulator.

Amplitudes of radial betatron oscillations of particles, according to (1), are increased in the second approximation (selectivity like (2) is not introduced) according to the law

$$\delta A_x^2 = (\delta x_\eta)^2 n, \quad A_x = \sqrt{A_{x,0}^2 + (\delta x_\eta)^2 n}, \qquad (5)$$

where $n = 1,2,3... n_{max} = 2\sigma_{\varepsilon,0}/\delta E = 2\sigma_{\eta,0}/\delta x_\eta$ is the number of interactions of particles with their URW, $\delta E = \overline{P} \cdot T$ is the energy loss of particles in kicker magnets, $T$ is the revolution period of particles in the storage ring, $\delta x_\eta = \dot{x}_{\eta,in} \cdot T$, $n \simeq t/T$. The betatron beam size $\sigma_x$ is changed by the same law:

$$\sigma_x = \sqrt{\sigma_{x,0}^2 + (\delta x_\eta)^2 n/2}, \qquad (6)$$

where $\sigma_{x,0}$ is the initial betatron beam size.

The expressions (4), (5) are valid until the distance between images of closed orbits of particles and the screen is larger then images of their amplitudes of betatron oscillations. If the distance is less, the URWs emitted by particles in the pick-up undulator under conditions of negative deviations $x_\beta < 0$ from their closed orbits will be absorbed by the screen and that is why will not lead to decrease of the energy of particles and to increase of their amplitudes of betatron oscillations in the kicker undulators. At the same time URW emitted by particles under conditions of positive deviations $x_\beta > 0$ will pass the screen and will lead to a decrease both their energy and amplitudes of betatron oscillations. If images of closed orbits of particles are deepened in the screen to the depth larger then images of their amplitudes of betatron oscillations the closed orbits will be stopped.

The non-exponential damping time in this scheme of cooling in the longitudinal direction is

$$\tau = \frac{2\sigma_{E,0}}{\overline{P}}. \qquad (7)$$

The damping time (7) is much less then damping time $\tau \simeq E/\overline{P}$ for schemes being described by Robinson's damping criterion [1].

The increase of the betatron beam size (6) for the damping time (7) determined by the term $(\delta x_\eta)^2 n/2 = \sigma_{\eta,0} \delta x_\eta$ can be neglected as usually the initial betatron size $\sigma_{x,0} \gg \sqrt{\delta x_\eta \sigma_{\eta,0}}$.

Six-dimensional phase space volume occupied by the beam (6D emittance) $\in_4 \sim \sigma_\varepsilon \sigma_x^2 \sigma_z^2$, according to (4), (6), is decreased up to zero in this non-exponential idealized process of cooling

The considered scheme of EOC is similar to radiative ion cooling one using broadband laser beams with sharp low frequency edges corresponding to excitation of ions of minimum and higher energies only and internal selectivity based on resonance ion properties [1], [4].

II. A screen overlap the URW beam in the image plane of the particle beam. A narrow slit in the screen is moving in the direction of the edge of the particle beam image corresponding to minimum particle energy with a velocity $v_{slit} < \dot{x}_{\eta,in}$ and open the way for a part of URW beam. This part of URWs is amplified. Particles emitted this part of URW beam interact with amplified URWs, decrease their energy similar to laser cooling of ion beams by monochromatic laser beam [5]. First, the slit open the way for URW emitted in the pick-up undulator by particles with higher energies and higher positive deviations $x_\beta > 0$ from their closed orbits. In this case a decrease both the energy and amplitudes of betatron oscillations of particles occur in kicker undulators (where $x_\beta < 0$).

III. A screen overlap the URW beam in the image plane of the particle beam and moves in the direction of the edge of the image corresponding to minimum particle energy with an image velocity corresponding to the closed orbit velocity $v_{scr} < \dot{x}_{\eta,in}$. First, the screen open the way for URW emitted in the pick-up undulator by particles with higher energies and higher positive deviations $x_\beta > 0$ from their closed orbits. In this case a decrease both the energy and amplitudes of betatron oscillations of particles occur in kicker undulators.

So the considered schemes of EOC II and III are going both in the longitudinal and transverse degrees of freedom. After the slit or screen will open images of all particles of the beam, the optical system must be switched off. Then the cooling process can be repeated.

## Geometry II

A pick-up undulator followed by even number of kicker undulators installed in straight sections of a storage ring. The distances between both pick-up and kicker neighboring undulators are determined by the phase advance equal to $(2p+1)\pi$. Particles enter kicker undulators at decelerating phases.

In this scheme, the deviations of particles in "$i$" and "$i+1$" undulators are $x_{\beta_i} = -x_{\beta_{i+1}}$ and that is why the decrease of energy of particles in undulators does not lead to change of the particle's betatron amplitudes at the exit of the last undulator. So it leads to cooling of the particle beam in the longitudinal plane only.

## Geometry III

A pick-up undulator and even number of kicker undulators are installed in straight sections of a storage ring at distances determined by a phase advance $(2p+1)\pi$ (like to the previous case) but optical paths are shifted to $\pi$ (contrary to the previous case). That is why particles decrease their energy in odd undulators and increase it in even ones.

In this case the change of the energy of particles in undulators leads to decrease of their betatron amplitudes and does not lead to change of their energy at the exit of the last undulator. In this case cooling of the particle beam is going in the transverse plane only.

The goal of usage of pick-up and kicker undulators, laser and optical systems in the scheme of EOC is similar to one in the method of optical stochastic cooling (OSC) [6]-[9]. Use of lenses and location of screens in the image plane is a principal moment for selection of particles in the method of EOC.

## TO THE THEORY OF EOC

Below we consider the theory of the EOC for the scheme 3 in the Geometry I.

The velocity of a particle closed orbit $\dot{x}_\eta$ depends on the probability $W$ of the particle to emit in the pick-up undulator such URWs which pass by the screen and interact with this particle in the kicker undulator. This probability is determined by the projection $A'$ of the particle betatron oscillations amplitude to the image plane and the distance $x_{scr} - x'_\eta$ between the edge of the screen and the projection to the image plane the particle closed orbit. Particles interact with their URW at every turn ($W = 1$) and the radial velocity of their closed orbit reaches a value $\dot{x}_{\eta,in} < 0$ if the image of their closed orbits come out the screen at a distance larger than images of amplitudes of their betatron oscillations. In the general case $\dot{x}_\eta = W \cdot \dot{x}_{\eta,in}$.

The value $W$ can be calculated by the next way (see [1], [2]). The deviation of a particle from its closed orbit at the location of the pick-up undulator takes on values $x_{\beta n} = A \cos \varphi_{\beta n}$, where $\varphi_{\beta n} = 2\pi \nu_x n + \varphi_{\beta 0}$, $n = 1, 2, 3, ...$ Different values $x_{\beta n}$ in the region $(-A, A)$ will occur with equal probability if $\nu_x \simeq p/q$,

$q \gg 1$ and tune is far from forbidden resonances, $p$ and $q$ are integers. The image of the particle pass by the screen and URW interact with the particle in the kicker undulator at every turn, if projected to the image plane deviation of a particle from its closed orbit $x'_{\beta n} \geq x_{scr} - x'_\eta$. This condition is valid if the particle phases $\varphi_{\beta n}$ are in the range $2\varphi_{scr}$, where $\varphi_{scr} = \arccos \xi$, $\xi = (x_{scr} - x'_\eta)/A'$. URWs of particles do not pass by the screen at the range of phases $2\pi - 2\varphi_{scr}$. It follows that the probability can be presented in the form $W = \varphi_{scr}/\pi$ and the value $\dot{x}_\eta = \varphi_{scr} \dot{x}_{\eta,in}/\pi$.

The behavior of amplitudes of betatron oscillations of particles is determined by (1). Particles cross the pick-up undulator at different deviations $x_\beta$ from their closed orbits in the range of phases $2\varphi_{scr}$. That is why the average rate of change of amplitudes $\partial A/\partial x_\eta = \bar{x}_{\beta 0}/A$, where $\bar{x}_{\beta 0} = (1/\varphi_{scr}) \int_0^{\varphi_{scr}} A \cdot \cos \varphi_{\beta n} d\varphi_{\beta n} = A \operatorname{sinc} \varphi_{scr}$ and $\operatorname{sinc} \varphi_{scr} = \sin \varphi_{scr}/\varphi_{scr}$.

Thus, the evolution of amplitudes and closed orbits or their images is determined by the system of equations

$$\frac{\partial A'}{\partial x'_\eta} = \frac{\partial A}{\partial x_\eta} = \operatorname{sinc} \varphi_{scr}, \quad (A' > 0)$$

$$\frac{1}{\dot{x}'_{\eta,in}} \frac{\partial x'_\eta}{\partial t} = \frac{1}{\dot{x}_{\eta,in}} \frac{\partial x_\eta}{\partial t} = \frac{\varphi_{scr}}{\pi}, \quad (\varphi_{scr} < \pi, A' > 0)$$

$$\frac{1}{\dot{x}'_{\eta,in}} \frac{\partial x'_\eta}{\partial t} = \frac{1}{\dot{x}_{\eta,in}} \frac{\partial x_\eta}{\partial t} = 1, (\varphi_{scr} = \pi, A' = 0) \quad (8)$$

where $\dot{x}'_{\eta,in}$ is the image of the velocity $\dot{x}_{\eta,in}$.

The position of the image of the closed orbit according to the definition of $\xi$, is determined by the relation $x'_\eta = x_{scr} - \xi A'(\xi)$. The time derivative of the image of the closed orbit is $\partial x'_\eta/\partial t = v_{scr} - [A' + \xi(\partial A'/\partial \xi)] \cdot \partial \xi/\partial t$ where $v_{scr} = dx_{scr}/dt$. Substituting this value to the second equation in (8) and using the condition $\dot{x}'_{\eta,in}/[A'(\xi) + \xi(\partial A'/\partial \xi)] = \dot{x}_{\eta,in}/[A(\xi) + \xi(\partial A/\partial \xi)]$ we obtain the derivative

$$\frac{\partial \xi}{\partial t} = \frac{\dot{x}_{\eta,in}}{\pi} \frac{\pi k_{scr} - \varphi_{scr}}{A(\xi) + \xi(\partial A/\partial \xi)}, \quad (9)$$

where $k_{scr} = v_{scr}/\dot{x}'_{\eta,in} > 0$ is the relative radial velocity of the screen. Using this equation we can transform the first equation in (8) to $\operatorname{sinc} \varphi_{scr}(\xi) = (\partial A/\partial \xi)(\partial \xi/\partial t)/(\partial x_\eta/\partial t) = (\pi k_{scr} - \varphi_{scr})(\partial A/\partial \xi)/[A + \xi(\partial A/\partial \xi)] \varphi_{scr}$ or $\partial \ln A/\partial \xi = \sin \varphi_{scr}/[\pi k_{scr} - (\varphi_{scr} + \xi \sin \varphi_{scr})]$.

The solution of this equation has the form

$$A = A_0 \exp \int_1^\xi \frac{\sin\varphi_{scr} d\xi}{\pi k_{scr} - (\varphi_{scr} + \xi \sin\varphi_{scr})} =$$
$$A_0 \exp \int_1^\xi \frac{\sqrt{1-\xi^2} d\xi}{\pi k_{scr} - \arccos\xi - \xi\sqrt{1-\xi^2}}. \quad (10)$$

We took into account that at the moment of the first interaction of a particle with the screen $t_{int}$ the amplitude $A(t_{int}) = A_0$, $\xi(t_{int}) = \xi_{int} = 1$.

In the method of EOC the fist term in the Eq. (8) $\sin\varphi_{scr} = \sqrt{1-\xi^2}$ and last term in the dominator of the Eq. (10) have opposite signs to similar terms in Eqs of the method of enhanced emittance exchange [1]. That is why the expression (10) keep integral form.

Substitution $A$ and $\partial A/\partial \xi$ in (9) leads to the time dependence $\xi(t)$ in the inverse form

$$t - t_{int} = \frac{\pi A_0}{\dot{x}_{\eta,in}} \psi(k_{scr}, \xi), \quad (11)$$

where

$$\psi(k_{scr}, \xi) = \int_\xi^1 \frac{d\xi}{\pi k_{scr} - \arccos\xi - \xi\sqrt{1-\xi^2}} \times$$
$$\exp \int_1^\xi \frac{\sqrt{1-\xi^2} d\xi}{\pi k_{scr} - \arccos\xi - \xi\sqrt{1-\xi^2}}.$$

The behavior of the amplitude (10) depends on the relative velocity of the screen $k_{scr}$.

If $k_{scr} < 1$, the denominator of (10) pass zero value at a parameter $\xi_c < 1$ determined by the equation

$$\pi k_{scr} - \arccos\xi_c - \xi_c\sqrt{1-\xi_c^2} = 0. \quad (12)$$

In this case the integral in (10) tends to $-\infty$ and the final amplitude tends to zero. After that the position of the image of the closed orbit follow the edge of the screen. At the end of the cicle of cooling all closed orbits are gathered at the edge of the screan. Radial and longitudinal phase space volumes occupied by the beam become zero.

If $k_{scr} > 1$, the integral in (10) determined by the condition $\xi = \xi_c = -1$ tends to negative finite limit and the final amplitude stay finite and decreased.

Closed orbits of particles having initial amplitudes of betatron oscillations $A_0$, according to (11), penetrate into the target to a depth greater than their final amplitudes of betatron oscillations $A_f$ at a moment

$$t_c = t_{int} + \frac{\pi A_0}{\dot{x}_{\eta,in}} \psi(k_{scr}, \xi_c). \quad (13)$$

The position of the image of the closed orbit, according to the definition of $\xi$ in the time interval $t_{int} < t < t_c$, can be presented in the form

$$x'_\eta = x_{scr,int} + v_{scr}(t - t_{int}) - \xi A'(\xi) =$$

$$x'_{\eta,0} - \Psi(k_{scr}) A'_0, \quad (14)$$

where $x_{scr,int}$ is the position of the screen at the moment $t_{int}$, $\Psi(k_{scr},\xi)=1+\pi k_{scr}\psi(k_{scr},\xi)-\xi A'(\xi)/A'_0$. We have used a condition $x'_{scr,int} = x'_{\eta,0} + A'_0$.

At the moment $t_c$ the position of the closed orbit of a particle is determined by the equation

$$x_{\eta,c} = x_{\eta,0} - \Psi(k_{scr}, \xi_c) A_0, \quad (15)$$

The moment $t_{int}$ depends on the initial conditions for particles $A_0$, $x_{\eta,0}$. If the edge of the target has a position $x_{scr,0}$ at a moment $t = 0$, then the target will contact a particle at the moment $t_{int} = -(x_{scr,0} - x_{scr,int})/v_{scr}$.

If $k_{scr} > 1$, closed orbits of particles are deepened into the target to a depth greater than amplitudes of betatron oscillation of particles at $t > t_c$. After that amplitudes of betatron oscillations of particles are constant, the velocity of their closed orbits is maximum and the orbits are moving according to the law $x_\eta(t) =$

$$x_{\eta,c} + \dot{x}_{\eta,in}(t - t_c) = x_{\eta,0} - \Psi(k_{scr})A_0 + \dot{x}_{\eta,in}[t - \pi A_0 \psi(k_{scr},\xi_c)/\dot{x}_{\eta,in} + (x_{scr,0} - x_{\eta,0} - A_0)/v_{scr}] \text{ or}$$

$$x_\eta(t) = x_{\eta,0}\frac{k_{scr}-1}{k_{scr}} - A_0[\Psi(k_{scr}) + \pi\psi(k_{scr},\xi_f)$$
$$+\frac{1}{k_{scr}}] + \dot{x}_{\eta in}(t + \frac{x_{scr,0}}{v_{scr}}) \quad (t > t_c). \quad (16)$$

The relative energy spread of the cooled beam, according to (16) is

$$\frac{\sigma_{E,f}}{\sigma_{E,0}} = \frac{\Delta x_\eta}{\Delta x_{\eta,0}} = \frac{k_{scr}-1}{k_{scr}} + R[\Psi(k_{scr}) - \pi\psi(k_{scr},\xi_c) + \frac{1}{k_{scr}}], \quad (17)$$

where $R = \sigma_{x,0}/\sigma_{\eta,0}$, $\sigma_{\eta,0}$ is the initial spread of closed orbits of the beam.

The duration of the cycle of the enhanced damping of the particle beam can be presented in the form $\tau = 2\sigma_{\eta,0}/v_{scr} + (t_c - t_{int})$ or

$$\tau = \frac{2\sigma_{\eta,0}}{k_{scr}|\dot{x}_{\eta,in}|} + \frac{\pi\sigma_{x,0}\psi(k_{scr},\xi_c)}{|\dot{x}_{\eta,in}|} =$$
$$\frac{2\sigma_{E,0}}{k_{scr}\bar{P}}\left[1 + \frac{\pi}{2}k_{scr}R\,\psi(k_{scr},\xi_c)\right] \quad (18)$$

The second term in (18) determine the damping time in the transverse plane.

According to (18), the higher the dispersion function of the storage ring at the location of the kicker undulator, the higher jumps $\delta x_\eta$ of the closed orbits of particles and $\dot{x}_\eta$, the higher the rates of betatron amplitudes decrease if $k_{scr} = const$. At the same time the lower the beta-function of the ring in this location, the lower $\sigma_{x,0}$, the higher the rate of betatron amplitudes decrease in the transverse direction as well. In this case the value $R$ and the cooling time (18) are decreased. Based on this observation, one can use combination of fast emittance exchange and EOC. In this case the energy spread $\sigma_{E,0}$ of the particle beam can be decreased $D \gg 1$ times and the radial betatron beam size will be increased $\sqrt{D}$ times in the process of the emittance exchange. Then EOC must be used. First of all such scheme can be applied to muon cooling.

The damping time of the particle beam in the method of EOC is much shorter than ordinary one when determined by the Robinson's damping criterion (proportional to the energy spread of the beam but not by the initial energy of particles). Moreover, in the approximation of neglecting of closed orbit jumps of particles in the process of their energy loss and diffraction processes considered below, the degree of EOC is much higher (zero at $k_{scr} < 1$) than ordinary $1/e$ reduction of the beam emittance for one damping time.

## THE RATE OF THE ENERGY LOSS

The total energy radiated by a relativistic particle traversing a given undulator magnetic field $B$ of finite length is given by

$$E_{tot} = \tfrac{2}{3} r_p^2 \overline{B^2} \gamma^2 M\lambda_u, \quad (19)$$

where $\overline{B^2}$ is an average square of magnetic field along the undulator length $M\lambda_u$, $\gamma$ is the relativistic factor, $M$ is the number of the undulator periods, $\lambda_u$ is the length of the undulator period, $r_p = Z^2 e^2 / M_p c^2 \simeq 1.53 \cdot 10^{-16} Z^2 / A$ is the classical radius of the particle. For a plane harmonic undulator, $\overline{B^2} = B_0^2/2$, where $B_0$ is the peak of the undulator field. For helical undulator $\overline{B^2} = B_0^2$.

The wavelength of the $k^{th}$ harmonic of the undulator radiation (UR)

$$\lambda_{UR,k} = \lambda_u (1 + K^2 + \vartheta^2)/2k\gamma^2, \quad (20)$$

where $\vartheta = \gamma\theta$; $\theta$, the angle between the closed orbit of the particle in the pick-up undulator and observation point; $K$ is the deflection parameter given by

$$K = \frac{Ze\sqrt{\overline{B^2}}\lambda_u}{2\pi M_p c^2}. \quad (21)$$

The number of the equivalent photons in the URW, emitted on the first harmonic, according to (19) – (21) and the condition $E_1 = E_{tot}/(1+K^2)$, becomes

$$N_{ph,1} = \frac{E_1}{\hbar\omega_{1,min}} = \tfrac{2}{3}\pi\alpha M Z^2 \frac{K^2}{1+K^2}, \quad (22)$$

where $\omega_{1,max} = 2\pi c/\lambda_{UR,1,min}$, $\lambda_{UR,1,min} = \lambda_{UR,1}|_{\theta=0}$.

In the regime of small deflection parameter $K < 1$, the spectrum of radiation emitted in the undulator with harmonic transverse magnetic field, is given by

$$dE_1/d\xi = E_1 f(\xi), \quad (23)$$

where $f(\xi) = 3\xi(1-2\xi+2\xi^2)$, $\xi = \lambda_{UR,1,min}/\lambda$, ($0 \leq \xi \leq 1$), $\int f(\xi)d\xi = 1$.

The bandwidth of the UR emitted at a given angle $\theta$ is

$$\frac{\Delta\omega}{\omega} = \frac{1}{kM}. \quad (24)$$

Below we accept a Gaussian distribution for the energy flow in the URWs, the image of the particle beam taken from the pick-up undulator is transformed by the optical system to centers of kicker undulators, the image coincide with the beam position at these undulators, $Z_R = M\lambda_u/2$. In this case the Rayleigh length of the URW $Z_R = 4\pi\sigma_{URW}^2/\lambda_{UR,1,min}$, the rms waist size $\sigma_{URW} = \sqrt{Z_R\lambda_{UR,1,min}/4\pi}$ and the rms electric field strength $\sqrt{\overline{E_{URW}^2}}$ of the non-amplified URW in the kicker undulator

$$\sqrt{\overline{E_{URW}^2}} = \sqrt{\frac{2E_{tot}}{M\lambda_{UR,1,min}\sigma_{URW}^2}} = \sqrt{\tfrac{8}{3}}\frac{r_p\sqrt{\overline{B^2}}\gamma^2}{(1+K^2)\sigma_{URW}} =$$

$$\sqrt{\frac{128\pi\overline{B^2}}{3M}}\frac{r_p\gamma^3}{\lambda_u(1+K^2)} = \sqrt{\frac{32\pi\overline{B^2}}{3M}}\frac{r_p\gamma}{\lambda_{UR,1,min}}. \quad (25)$$

The rate of the energy loss for particles in the fields of kicker undulates and amplified URWs reach maximum

$$\overline{P} = eZKM\lambda_u f N_{kick}\gamma^{-1}\sqrt{\alpha_{ampl}\overline{E_{URW}^2}} =$$

$$\frac{8\pi r_p f N_{kick}\sqrt{eZLM_pc^2\alpha_{ampl}\overline{B^2}}}{\sqrt{3}\,\lambda_{UR,1,min}}K^{3/2} \quad (26)$$

if particles enter kicker undulator at decelerating phase corresponding to maximum of the electric field strength. Here $f$ is the revolution frequency, $N_{kick}$ is the number of kicker undulators, $L = M\lambda_u$ is the length of undulators, $\alpha_{ampl}$ is the gain in optical amplifier. We used the expressions $M = L(1+K^2)/2\gamma^2\lambda_{UR,1,min}$, $\gamma = \sqrt{\pi M_p c^2 K(1+K^2)/eZ\sqrt{\overline{B^2}}\lambda_{UR,1,min}}$ as well.

## REGIMES OF COOLING

Wavelets of UR emitted by a particle in the pick-up undulator and amplified in an optical amplifier interact efficiently with the same particle in the kicker undulators. Amplified radiation from one particle does not disturb trajectories of other particles if an average distance between particles in a longitudinal direction is more, than the length of the URW, $M\lambda_{UR,1,min}$. In this case the beam current $i < i_c$, where

$$i_c = \frac{Zec}{M\lambda_{UR}} = \frac{4.8\cdot 10^{-9}Z}{M\lambda_{UR}}[A]. \quad (27)$$

is a characteristic current.

URWs of particles in kicker undulators at $i > i_c$ can be overlapped partially both in longitudinal and transverse planes. The degree of overlapping is determined by a parameter $n_s$ named "number of particles in an URW sample". The value $n_s = (i/i_c)(s_{URW}/s_{b,f})$ if $s_{b,f}/s_{URW} > 1$ and $n_s = i/i_c$ if $s_{b,f}/s_{URW} < 1$, where $s_{b,f}$ is the transverse area of the particle beam in the kicker undulator at the end of cooling cycle, $s_{URW} = 2\pi\sigma_{URW}^2$ is the effective transverse area of the URWs.

In the regime $n_s \gg 1$ the interaction of the URW emitted by a particle $k$ in the pick-up undulator leads to a decrease of the energy and/or the amplitude of betatron oscillations of the particle. At the same time $n_c - 1$ URWs emitted by other particles in the same URW sample have random phases for the particle $k$. They do not lead to a change of the energy and amplitude of betatron oscillations of the particle $k$ in average in the first approximation and will lead to change its energy and amplitude in the second one (a particle $k$ has equal probability to enter kicker undulators at positive and negative phases in URWs emitted by other particles).

The electric field strength and energy jumps of particles in $n_s - 1$ URWs produced by other particles and in noise photons in kicker undulators are proportional to $\sqrt{n_s - 1 + n_n}$ in average, where $n_n$ is the ratio of the number of noise photons at the amplifier front end to the number of signal photons. Phases of the particle $k$ in these URWs are random and that is why jumps of their closed orbits are distributed in the range $\pm \delta x_\eta'' = \pm \delta x_\eta \sqrt{n_s - 1 + n_n}$.

The closed orbit position uncertainty growth of a particle $k$ in kicker undulators produced by URWs of other particles, according to the binomial distribution of probability, is determined by the law $\sqrt{\overline{x_\eta^2}} = \delta x_\eta'' \cdot \sqrt{n}$, where $n$ is the number of interactions of the patricle with these URWs. In this case the rate of growth of the spread of the closed orbits is determined by the equation $d\overline{x_\eta^2}/dn = (\delta x_\eta'')^2$. The rate of damping of the spread is determined by the equation $d\sqrt{\overline{x_\eta^2}}/dn = \delta x_\eta$ or $d\overline{x_\eta^2}/dn = 2\delta x_\eta \sqrt{\overline{x_\eta^2}}$. The change of the spread of the closed orbits is determined by the equation

$$\frac{d\sqrt{\overline{x_\eta^2}}}{dn} = 2\delta x_\eta \sqrt{\overline{x_\eta^2}} + (\delta x_\eta)^2(n_s - 1 + n_n). \quad (28)$$

The solution of this equation can be presented in the form

$$n = -\frac{1}{\delta x_\eta}(\overline{x_\eta^2} - \overline{x_{\eta,0}^2}) +$$

$$\frac{1}{2}(n_s - 1 + n_n)\ln \frac{\sqrt{\overline{x_\eta^2}} + \frac{1}{2}\delta x_\eta(n_s - 1 + n_n)}{\sqrt{\overline{x_{\eta,0}^2}} + \frac{1}{2}\delta x_\eta(n_s - 1 + n_n)}. \quad (29)$$

In the limit $n \to \infty$ the spread of the closed orbits tend to

$$\left(\sqrt{\overline{x_\eta^2}}\right)_{eq} = \frac{1}{2}|\delta x_\eta|(n_s - 1 + n_n). \quad (30)$$

The rms betatron beam dimension growth is determined by the last term in (1) with replacement $(\delta x_\eta)^2 \to (\delta x_\eta'')^2$: $\overline{A_x^2} = (\delta x_\eta'')^2 n$. At the same time the rate of damping of the amplitude per one interaction is $dA_x/dn = d\sqrt{\overline{x_\eta^2}}/dn = \delta x_\eta$. It follows that the amplitude of betatron oscillations of particles in the beam is changed by the same law (29). In the limit $n \to \infty$:

$$\left(\sqrt{\overline{A_x^2}}\right)_{eq} = \left(\sqrt{\overline{x_\eta^2}}\right)_{eq} = \frac{1}{2}|\delta x_\eta|(n_s - 1 + n_n). \quad (31)$$

In the regime $i < i_c$, ($n_s = 1$) and if the noise of amplifier is neglected ($n_n = 0$), the limiting energy spread is determined by a jump of the closed orbit of particles $\delta x_\eta$.

## LATTICE LIMITATIONS

In the smooth approximation the relative phase shifts of particles in their URWs radiated in the pick-up undulator and displaced to the entrance of kick undulators depend on theirs energy and amplitude of betatron oscillations. If we assume that the longitudinal shifts of URWs $\Delta l < \lambda_{UR,\min}/2$, then the amplitudes of betatron oscillations, transverse horizontal emittance of the beam and the energy spread of the beam, in the smooth approximation, must not exceed the values

$$A_{x,\lim} = \frac{\sqrt{\lambda_{UR}\lambda_{x,bet}}}{\pi}, \qquad \varepsilon_{x,\lim} < 2\lambda_{UR}/\pi,$$

$$(\frac{\Delta\gamma}{\gamma})_{\lim} = \frac{\beta^2}{\eta_c}\frac{\lambda_{UR}}{2C}, \quad (32)$$

where $\lambda_{x,bet} = C/v_x$, C is the circumference of the ring, $\eta_c = \alpha_c - \gamma^{-2}$ and $\alpha_c$ are local slip and momentum compaction factors between undulators.

Strong limitations (32) to the energy spread can be overcame if, special elements in storage ring lattices (short inverted dipoles, quadrupole lenses et al.) to decrease the slip [10]-[13] will be used along the paths between undulators. A change in time of optical paths of URWs can be produced according to the decrease of the high energy edge of the being cooled beam as well. With cooling of fraction of the beam at a time only, the lengthening problem diminishes also as $\Delta E/E$ now stands for the energy spread in the part of the beam which is under cooling at the moment.

## AMPLIFIER

The power of the amplifier is determined by the power of the amplified URWs

$$P_{ampl} = \varepsilon_{URW} \cdot f \cdot N_p, \quad (33)$$

where $\varepsilon_{URW} = \hbar\omega_{1,\max}N_{ph,1}\alpha_{ampl}$ is the energy in the amplified URW; $N_p$, the number of particles in the ring. Different amplifiers can be used by analogy with the case of OSC [6]–[9]. Parametric amplifiers can be used as well [14]-[16].

## SELECTIVITY TECHNIQUE

The transverse selectivity of radiation by movable screen can be arranged with help of electro-optical elements. These elements contain crystals, which change its refraction index while external voltage applied. This technique is well known in optics [17]. In simplest case the sequence of electro-optical deflector and a diaphragm followed by optical lenses, allow controllable selection of radiation generated by different parts of the beam.

## LIMITING RESOLUTION OF THE OPTICAL SYSTEM

UR emitted by a particle in the undulator installed in a stright section of a storage ring propagates in a narrow cone of a solid angle $\theta_c \simeq 1/\gamma$ to the particle velocity. The limiting resolution angle of any optical system is determined by the equation $\Delta\theta_r = 1.22\lambda/d$, where $\lambda = \lambda_{UR}$, $d$ is the diameter of the UR light on the first mirror. The value $d \simeq l\theta_c$, where $l$ is the distance from the pick-up undulator to the first mirror. That is why the space resolution of the particle beam is limited by the value

$$\delta x = l\Delta\theta_r \simeq 1.22\lambda\gamma. \qquad (34)$$

The resolution (34) determines the limiting degree of cooling of particle beams in storage rings. To produce a high degree of cooling the initial radial beam dimension at pick-up undulator $\sigma_{b,0} = \sqrt{\sigma_{x,0}^2 + \sigma_{\eta,0}^2}$ must be much bigger $\delta x$. To increase the degree of cooling we can increase the beam dimension $\sigma_{b,0}$ by increasing the dispersion- and beta-functions of the ring at the location of the pick-up undulator and by using small wavelength optical amplifier.

## COOLNG IN A BUCKET

Cooling of particles can be produced in the radio frequency (RF) bucket. In this case the screen must be moved to the position of the image of the equilibrium orbit and stopped at this position. Cooling cycles must be repeated periodically.

## EXAMPLE

Enhanced optical cooling of fully stripped $^{207}_{82}Pb$ ion beam in the CERN LHC at the injection energy [18]. The relevant parameters of the LHC:

circumference $C \simeq 26.66$ km,
bending radius $R \simeq 2804$ m,
revolution frequency $f = 1.12 \times 10^4$ Hz,
horizontal tune at injection $v_x = 64.28$,
momentum compaction $\alpha_c = 3.23 \cdot 10^{-4}$,
slip factor $\eta_c = 3.18 \cdot 10^{-4}$,
relativistic factor $\gamma = 190.5$,
particle energy $M_p c^2 \gamma = 36.9$ TeV,
energy spread $\Delta\gamma/\gamma = 3.9 \cdot 10^{-4}$,
transverse normalized emittance $\epsilon_{x,n} = 1.4\ \mu m$,
total number of particles $N_{Pb} = 4.1 \cdot 10^{10}$,
betatron oscillation length $\lambda_{x,bet} = 414.7$ m,
beta and dispersion functions at kicker undulators:
$\beta_x = 25.0$ m, $\eta_x = 2.0$ m,
betatron beam size at pick-up undul. $\sigma_{x,0} = 0.43$ mm,
dispersion beam size $\sigma_{\eta,0} = 0.95$ mm,
total beam size $\sigma_{b,0} = 1.1$ mm.

One pick-up and 10 kick undulators are used. They have parameters:

undulator period $\lambda_u = 4$ cm,
number uf undulator periods $M=300$,
rms magnetic field $\sqrt{B^2} = 10^5$ Gs,
the relative radial velocity of the screen $k_{scr} = 0.5$,
function $\psi(0.5,1.0) = 1.37$,
amplifier gain goes to be $\alpha_{ampl} = 10^6$.

At these parameters:

min. undulator wavelength $\lambda_{UR,1,min} = 5.5 \cdot 10^{-5}$ cm,
photon energy $\hbar\omega_{1,max} = 2.25$ eV,
rms waist size at kicker undulator $\sigma_{URW} = 0.51$ mm,
number of the emitted photons $N_{ph,1} \simeq 2.0$,
characteristic current $i_c = 0.024$ mA,
character. number of particles $N_c = i_c/e\,f = 1.7 \cdot 10^8$,
parameter $n_s = i/i_c \simeq 241$,
parameter $n_n = 0.5$
(if one noise photon is at the amplifier front end),
deflecting parameter $K=0.0081$,
electric field strength $\sqrt{E_{URW}^2} \cong 0.172$ V/cm,
power of the amplified URWs $\bar{P} = 8.06 \cdot 10^7$ eV/sec,
particle energy loss $\Delta E_{loss} = \bar{P}/f = 7.19 \cdot 10^3$ eV/rev,
change of particle orbit position $\delta x_\eta = 3.9 \cdot 10^{-8}$ cm,
energy spread of the beam $M_p c^2 \Delta\gamma = 2\sigma_{\eta,0} =$
$= 1.44 \cdot 10^{10}$ eV,
energy in the amplified URW $\varepsilon_{URW} = 7.2 \cdot 10^{-12}$ J,
beam dimensions $\left(\sqrt{A_x^2}\right)_{eq} = \left(\sqrt{x_\eta^2}\right)_{eq} = 9.4$ mkm,
power of amplifier $P_{ampl} = 331$ W,
limiting amplitude $A_{x,lim} = 0.48$ cm,
limiting energy spread $(\frac{\Delta\gamma}{\gamma})_{lim} = 6.49 \cdot 10^{-8}$,
damping time $\tau = 8.81$ min,

In this example the relative energy spread of the beam $(\Delta\gamma/\gamma)_{lim} \ll \Delta\gamma/\gamma \ll \Delta\omega/\omega = 1/M$, transverse beam dimensions $\sigma_{x,0} < A_{x,lim}$. It means that there is no problem with the dependence of phase shifts of particles on their amplitudes. At the same time there is a necessity in special elements in storage ring lattices to decrease the slip. The space resolution of the particle beam is limited by the value $\delta x = 0.128$ mm ($\delta x \ll \sigma_{b,0}$). This resolution does not permit to reach the evaluated above very small equilibrium beam dimensions. It can be increased by increasing the dispersion and beta functions in the pick-up undulators.

Damping time of a particle beam, according to (18), (26), $\tau \sim \sigma_{E,0} \lambda_{UR,1,\min} / K^{3/2} \sqrt{L \sqrt{\overline{B^2}}}$. If the length of the undulator $L = M\lambda_u$, the wavelength of the first harmonic of UR $\lambda_{UR,1,\min}$, the initial energy spread $\sigma_{\varepsilon,0}$ and $\overline{B^2}$ are constant, the damping time $\sim \sigma_{\varepsilon,0} / K^{3/2} |_{K \ll 1} \sim \sigma_{\varepsilon,0} / \gamma^3$.

We considered an example of EOC in LHC at the injection energy. Cooling under above conditions at the relative energy $\gamma = 973$, ($\lambda_u = 100$ cm, $K=0.2$) will lead to the damping time 4.08 sec.

## CONCLUSION

We considered EOC of particle beams in storage rings mainly for unbunched beam. The rate of EOC defined by the ratio of the energy spread to the rate of the energy loss is more than the rate of OSC. Methods of calculations developed in this paper can be useful for OSC as well (see Appendix).

The damping time (18) is proportional to both energy spread and radial betatron beam dimension. First cooling cycle can be long and include cooling of halo of the beam. Next cooling cycles can be much shorter to support the beam dimensions small.

The considered schemes of EOC are of great interest for cooling of fully stripped ion, proton and muon beams. Laser cooling, based on nuclear transitions has problems with low-lying levels [19]. EOC of heavy ions, on the level with OSC, is the most efficient. In this case the number of emitted photns $\sim Z^2$, where $Z$ is the atomic number [20].

## Appendix

Below we consider the scheme of OSC based on a quadrupole pick-up undulator and $N_{kick}$ ordinary kicker undulators installed in straight sections of a storage ring at a distance determined by a betatron phase advance for the lattice segment $(2p+1)\pi$. Pick-up and kicker undulators have equal numbers of periods. Particles enter kicker undulators at decelerating phases of URW.

The deflection parameter (21) and the rms electric field strength of the non-amplified URW emitted in the quadrupole undulator depend on the coordinate $x = x_\eta + x_\beta$ of a particle crossing the undulator by the law:

$$K_Q = (Ze\sqrt{\overline{G_x^2}} \lambda_{u,Q} / 2\pi M_p c^2) |x|, \quad \sqrt{\overline{E_{QURW}^2}} = \sqrt{32\pi \overline{G_x^2}/3M} (r_p \gamma / \lambda_{UR,1,\min}) |x|,$$

where $\lambda_{u,Q}$ is the period of the quadrupole undulator, $G_x$ is the gradient of the magnetic field strength of the undulator, coordinate $x = 0$ corresponds to the undulator axis.

The rate of the energy loss for a particle in kicker undulators is determined by the equation

$$\overline{P} = eZK_K M \lambda_{u,K} f N_{kick} \gamma^{-1} \sqrt{\alpha_{ampl} \overline{E_{QURW}^2}} (x/|x|) = eZr_p \cdot K_K N_{kick} f (\lambda_{u,K}/\lambda_{UR,1,\min}) \sqrt{32\pi \alpha_{ampl} M \overline{G_x^2}/3} \cdot x \quad (35)$$

where $K_K$ is the deflection parameter of the kicker undulators, $\lambda_{u,K}$ is the period of the kicker undulator. In the last term in Eq. (35) we took into account that in the quadrupole undulator the phase of the emitted URW is changed by $\pi$ if coordinate $x$ change sign.

The wavelength of the emitted UR (20) in this scheme depends on the parameter $K_Q$. To keep resonance between URW and particles in kicker undulator the ratio $\lambda_{u,Q}(1+K_Q^2) = \lambda_{u,K}(1+K_K^2)$ must be fulfilled. It means that if periods $\lambda_{u,Q}$ and $\lambda_K$ are time independent and the coordinate $x$ is decreased in the cooling process, the parameter $K_K$ must be changed during this process. The increase of the spread of the closed orbits can be done by increasing with time the dispersion function $\eta_x$ of the ring in the location of undulators [6]. The range of change of the parameter $K_K$ and dispersion function is lesser, if the initial parameter $K_Q \ll 1$. But this regime can be far from optimal one. The resonance condition must be applied to particles with higher energies. Another particles will have much smaller rate of the energy change if they are far from the resonance.

The change of the closed orbit position of a particle per its revolution period is $\delta x_\eta = \eta_x \beta^{-2} (\overline{P} T/E)$ or

$$\delta x_\eta = -G_r \cdot (x_\eta + x_\beta) \quad (36)$$

where

$G_r = 4eZr_p K_K N_{kick} \eta_x \lambda_{u,K} \sqrt{2\pi \alpha_{ampl} M \overline{G_x^2}} / \sqrt{3} \beta^2 E \cdot \lambda_{UR,1,\min}$. The value $x_\beta$ is a fast oscillating function. Average value $\overline{x_\beta} = 0$. In the approximation $K_K \simeq$ const $\ll 1$, $\eta_x$ =const, the average rate of the closed orbit position change is

$$\frac{\partial \overline{x_\eta}}{\partial t} = \frac{\delta x_\eta}{T} = -\frac{G_r}{T} \cdot \overline{x_\eta} \quad (37)$$

and its solution

$$x_\eta = x_{\eta,0} \cdot e^{-G_r \cdot t/T}. \quad (38)$$

The change of the betatron amplitude of a particle per its revolution period is $dA = (x_\beta/A)\delta x_\eta$ or, according to (36), $dA = -G_r \cdot x_\beta (x_\eta + x_\beta)/A$. The average rate of the betatron amplitude of a particle change is

$$\frac{\partial A}{\partial t} = -\frac{G_r}{A \cdot T} \cdot \overline{x_\beta^2} = \frac{G_r}{2 \cdot T} A. \quad (39)$$

The solution of this equation is

$$A = A_0 \cdot e^{-G_r \cdot t/2T}. \quad (40)$$

Let us compare the rate of OSC and EOC using the considered above example of ion cooling in LHC.

1. Maximal magnetic field of the pick-up undulator (both quadrupole and ordinary) do not exceed $10^5$ Gs.

For quadrupole undulator such field ($G_x \cdot D_x = 10^5$ Gs) is produced at the orbits positions $x_\eta = D_x = 1$ cm, where $2 D_x$ is the aperture of the storage ring ($D_x \simeq 10 \cdot \sigma_{b,0}$). That is why the initial rate of change of the closed orbit position $\delta x_\eta = -G_r \cdot \sigma_{x,0}$ for OSC is more then 10 times less. Moreover the rate for OSC is decreased (if dispersion function does not increased on time) and for EOC it stay constant. The amplitude of betatron oscillations of particles decay according to exponential law for OSC with small initial rate of decreese and for EOC it decay near to zero value for the damping time.

Similar situation (exponential decay) is for the scheme of OSC based on ordinary pick-up and kicker undulators considered in [7]. In this case the energy change of particles depends on the phase of their own URWs amplified in the optical amplifier at which the particles enter the kicker undulator. This scheme has an advantage in comparison with one based on quadrupole undulator as in this case it use ordinary pick-up undulator and that is why the value of the undulator magnetic field does not depend on radial coordinate and can be high ($10^5$ Gs).